\documentclass[11pt]{article}
\usepackage{blois2002,epsfig}

\def\be{\begin{equation}}
\def\ee{\end{equation}}
\def\bea{\begin{eqnarray}}
\def\eea{\end{eqnarray}}

\newcommand{\pbar}{$\overline{\pfont p}$}

\newcommand{\pbarhelium}{\pbar He${}^+$}
\newcommand{\pbarheliumion}{\pbar He${}^{++}$}
\newcommand{\pbhe}{\pbarhelium\ }

\newcommand{\us}{$\mu$s}

\newcommand{\pbarHethree}{\pbar$\,{}^3${He}${}^+$}
\newcommand{\pbarHefour}{\pbar$\,{}^4${He}${}^+$}
\newcommand{\el}{\ifmmode l\else$l$\fi}
\newcommand{\litter}{\ifmmode\ell\else$\ell$\fi}

\newcommand{\nuHFp}{$\nu_{\mathrm HF}^+$ }
\newcommand{\nuHFm}{$\nu_{\mathrm HF}^-$ }

\newcommand{\mupbar}{$\mu_{\overline{\rm p}}$ }

\newcommand{\LP}{$L_{\overline{\mathrm p}}$ }
\newcommand{\SP}{$S_{\overline{\mathrm p}}$ }
\newcommand{\Se}{$\vec{S}_e$}
\newcommand{\Lp}{$\vec{L}_{\overline{p}}$}
\newcommand{\Sp}{$\vec{S}_{\overline{p}}$}

\newcommand{\pfont}{\rm}

\begin{document}
\vspace*{4cm}
\title{HIGH PRECISION LASER AND MICROWAVE SPECTROSCOPY OF ANTIPROTONIC HELIUM}

\author{ E. WIDMANN \\
    for the ASACUSA collaboration}

\address{Department of Physics, University of Tokyo, 7-3-1 Hongo,
    Bunkyo-ku, Tokyo 113-0011, Japan}

\maketitle\abstracts{ This talk gives an overview of the recent results
on the precision spectroscopy of antiprotonic helium which was
performed by the ASACUSA collaboration at the Antiproton Decelerator of
CERN. The laser spectroscopy of energy levels of the antiproton has
reached a relative accuracy of $\sim 10^{-7}$, and by comparing the
experimental value for the transition wavelengths with theoretical
calculations, a CPT test on the equality of proton and antiproton
charge and mass of $<6\times10^{-8}$ has been obtained. In a recent
experiment, the hyperfine structure of the $(n,l)=(37,35)$ state of
antiprotonic helium has been measured for the first time with an
accuracy of $3 \times 10^{-5}$. }

\section{Antiprotonic helium}

The precision spectroscopy of antiprotonic helium is being performed by
the ASACUSA collaboration \cite{ASACUSA:97+00} at the Antiproton
Decelerator (AD) facility of CERN in Geneva, Switzerland. The
experiments constitute an extension of those performed previously by
the PS205 collaboration at
LEAR\cite{Iwasaki:91,Yamazaki:93,Yamazaki:01}. Here, highly excited
states of the neutral three-body system \pbar--He$^{2+}$--e$^- \equiv$
\pbarhelium\ (cf. Fig.~1) are investigated by means of laser and
microwave spectroscopy.
These states are formed when antiprotons are stopped in helium (in the
current experiments, helium gas of about 6 K and $0.1-2$ bar).
Antiprotons are captured by replacing one of the two ground-state
electrons of helium and therefore occupy states with principal quantum
numbers around $n_0=\sqrt{M^*/m}$ (=38.3 for \pbarHefour), where $M^*$
is the reduced mass of the \pbar--He$^{2+}$ system, and $m$ the
electron mass. Fig.~1 shows the energy level diagram of \pbarHefour,
which is divided in a metastable zone (solid lines) and a short-lived
one (wavy lines). For the metastable levels the Auger transition rate
is much smaller than the radiative transition rate, leading to
lifetimes of $\tau \sim$ \us, while for the short-lived ones the Auger
rate is much higher than the radiative one, resulting in lifetimes of
$\tau <$ 10 ns. Antiprotons initially captured around $n_0$ undergo
radiative transitions following cascades with $\Delta v \equiv
\Delta(n-l-1)=0$ ($n$ = principle and $l$ = angular momentum quantum
number) until they reach a short-lived state from which they ionize via
Auger transitions. The \pbarheliumion\ ion is then rapidly destroyed by
Starck-mixing in the dense surrounding helium medium.

This picture was proven to be correct by a series of laser spectroscopy
experiments at LEAR (\cite{Morita:94} and references in
\cite{Yamazaki:01}). A laser pulse tuned to the transition at the end
of a cascade will deexcite the antiprotons from the metastable to the
short-lived state, thus forcing them to immediately annihilate. Using
this signature, a total of 10 resonant transitions in \pbarHefour\ and
3 in \pbarHethree\ could be found at LEAR, thus confirming for the
first time experimentally the long-held belief that exotic particles
are initially captured around $n_0=\sqrt{M^*/m}$.

\begin{figure}[t]
\begin{center}
    \epsfig{file=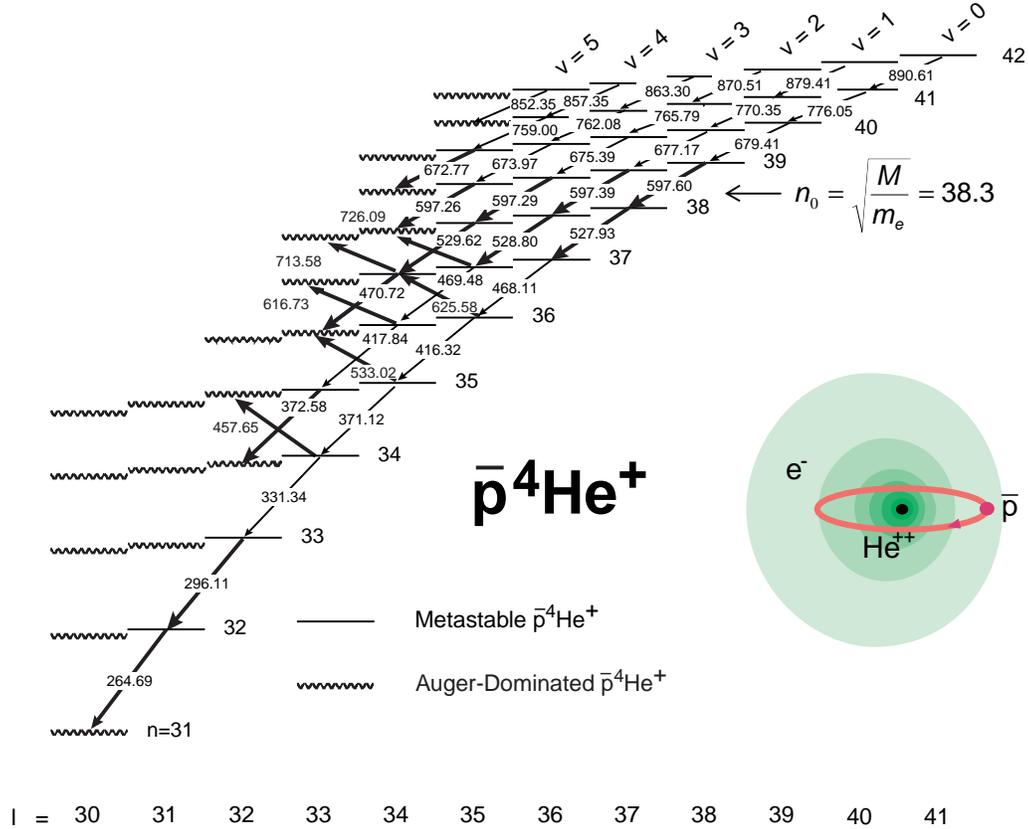,width=13.5cm}
    \caption{Right: structure of antiprotonic helium. Left: Energy
    level diagram of \pbarHefour. Metastable levels ($\tau\ge$ \us) are
    denoted by solid lines, Auger-dominated short-lived ones ($\tau \le
    10$ ns) by wavy lines. Arrows symbolize radiative transitions
    following the propensity rule $\Delta v =0$. Bold arrows correspond
    to experimentally observed transitions. They include so-called
    unfavoured ones with $\Delta v =2$. The transition wavelengths are
    denoted in units of nm.}
\end{center}
\end{figure}

\section{Laser spectroscopy of antiprotonic helium and CPT test of the
    antiproton charge and mass}

The AD is in operation since the year 2000. It provides pulses of $\sim
200$ ns length containing $2-4\times10^7$ antiprotons of 100 MeV/c
momentum (5.3 MeV kinetic energy). The distance between pulses is about
100 seconds. For the experiments at the AD a completely new
experimental setup was designed and installed, adopted to the pulsed
structure of the antiproton beam with very low repetition rate. Laser
spectroscopy of \pbhe was started from the begin of operation of the
AD, and many new transitions in \pbarHefour\ and  \pbarHethree\ were
found. This allows for a much more systematic study of state properties
and theoretical calculations.

\begin{figure}
    \begin{minipage}{8cm}
        \begin{center}
        \epsfig{file=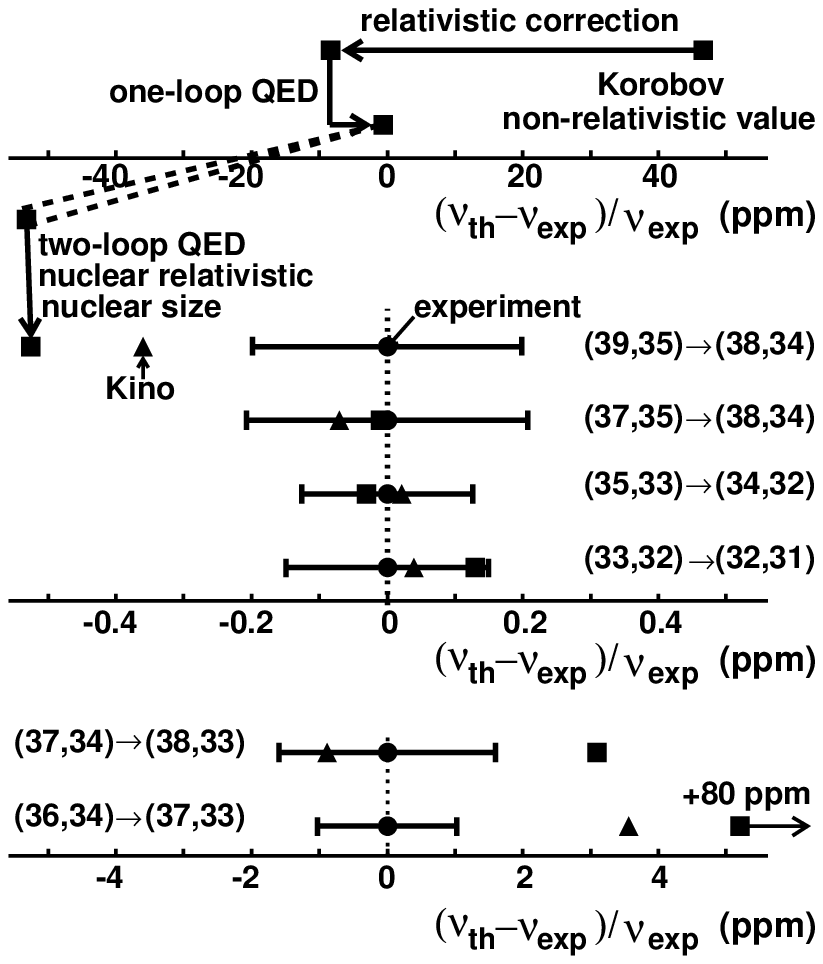,width=6cm}
        \end{center}
    \end{minipage}
    \begin{minipage}{8cm}
        \begin{center}
        \epsfig{file=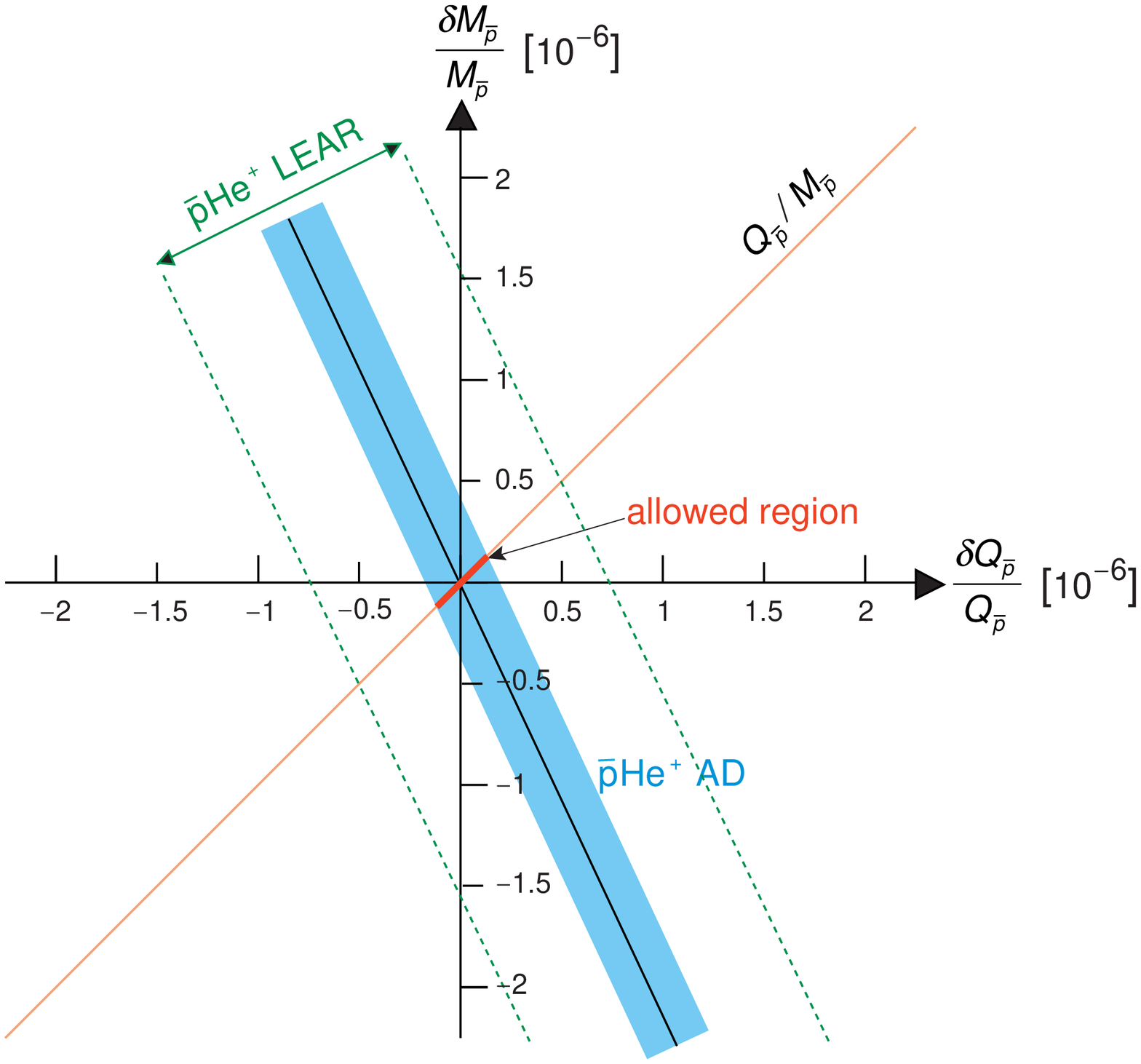,width=7cm}
        \end{center}
    \end{minipage}
    \caption{\label{fig:Widmann-th-exp} Left: Comparisons between
    experimental (filled circles with errors) and theoretical (squares
    \protect\cite{Korobov:00} and triangles
    \protect\cite{Kino:99,Kino:01}) values for six transition
    frequencies in \pbhe atoms. The upper four ``narrow'' (see text)
    transitions show agreement between theory and experiment to better
    than $5\times10^{-7}$. Right: Two-dimensional constraint on  the relative
    deviation of the proton and antiproton charge and mass obtained from
    the cyclotron frequency of the
    antiproton \protect\cite{Gabrielse:99} and from the
    spectroscopic studies of \pbarhelium\ at LEAR \protect\cite{Torii:99}
    and at the AD \protect\cite{Hori:01}. }

\end{figure}

At LEAR, we had succeeded in measuring the transition wavelength for
one transition with the best accuracy of 0.5 ppm ($5\times10^{-7}$)
\cite{Torii:99}. In order to improve this, we used a new laser system
and more sophisticated calibration of the wavelength measuring device
against an iodine reference cell \cite{Hori:01}. In the first
experiment, we also extended the measurements to six transitions
including two newly found ones in the UV region
($(n,l)=(35,33)\rightarrow(34,32)$ at 372 nm and
$(33,32)\rightarrow(32,31)$ at 296 nm) where theoretical calculations
are supposed to be most accurate. Two transitions had very large widths
($\Gamma > 15$ GHz) because of the short lifetime of their daughter
states, which makes both the experimental and theoretical accuracy very
low. For the other four ``narrow'' transitions ($\Gamma <50$ MHz), the
experimental error was reduced by a factor of four to
$1.3\times10^{-7}$, and all four transition wavelengths agreed with the
most sophisticated three-body calculations by Korobov \cite{Korobov:00}
and Kino et al. \cite{Kino:99,Kino:01} to better than $5\times10^{-7}$
(cf. Fig.~\ref{fig:Widmann-th-exp}) \cite{Hori:01}. At this level of
precision the calculations need to take into account relativistic
corrections, the Lamb shift and QED corrections up to order $\alpha^4$,
so that this experiment constitutes a stringent test of the validity of
three-body bound state QED calculations.

On the other hand, the agreement between experiment and theory can be
used to perform a CPT test on the proton/antiproton charge and mass,
since the theorists use in their calculations the numerically better
known mass and charge of the proton. The energy levels itself are
governed by the Rydberg constant $Ry \sim M_{\overline{\mathrm p}}
Q_{\overline{\mathrm p}}^2$. Taking into account the fact that the
antiproton cyclotron frequency $\omega_{\overline{\mathrm p}}\sim
Q_{\overline{\mathrm p}}/ M_{\overline{\mathrm p}}$ is known
\cite{Gabrielse:99} to be equal to the proton one to better than 1 in
$10^{10}$, the agreement of theory and experiment for the transition
energies in antiprotonic helium results in an upper limit for the
equality of proton and antiproton charge and mass of $6\times10^{-8}$
(90\% confidence level) \cite{Hori:01}, a factor of 8 better than our
own results at LEAR \cite{Torii:99} and a factor of 300 better than
previous measurements using X-rays of heavy antiprotonic atoms
\cite{PDG:00}.

In 2001 we went one step ahead and used an Radio Frequency Quadrupole
Decelerator (RFQD \cite{RFQD:design}) to further decelerate the
antiprotons to $\sim 60$ keV. These ultra-low energy antiprotons can be
stopped in helium gas of 30 K and 0.8 mbar, about a factor 1000 lower
density than usual. This allows to essentially eliminate the systematic
error coming from the extrapolation of the transition wavelength to
zero density necessary in the previous experiments. Furthermore, we
extended the number of transitions investigated to 7 in \pbarHefour\
and, for the first time, included 6 transitions of \pbarHethree.
Preliminary analysis shows that we can expect a further improvement of
a factor $3\ldots4$ for the experimental accuracy after analyzing all
new data.

\section{Measurement of the hyperfine structure of antiprotonic helium}

\begin{figure}[b]
        \begin{center}
        \epsfig{file=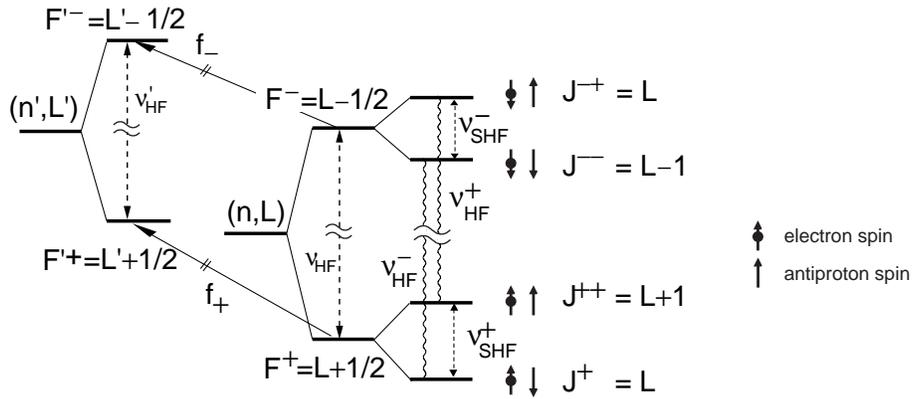,width=12cm}
        \end{center}
    \caption{\label{fig:HFS} Schematic view of the splitting of a
    \pbhe state and observable laser transitions from the $F^{\pm}$
    levels of a $(n,L)$ state to a daughter state $(n',L')$ (arrows).
    Wavy lines denote allowed magnetic transitions associated with an
    electron spin flip. }

\end{figure}

The main highlight of the 2001 run was the first observation of the
hyperfine splitting of the $(37,35)$ state using a
laser-microwave-laser triple resonance method \cite{Widmann:02}. In
this case ``hyperfine'' structure means the level splitting arising
from the interaction of the magnetic moments of the constituents of
\pbarhelium. The uniqueness of the resulting structure comes from the
fact that the antiproton in metastable states carries a very large
angular momentum of \LP$=33\ldots39$, which leads through the
interaction $\vec{L_{\overline{\mathrm p}}}\cdot \vec{S_e}$ with the
electron spin $S_e$ to the dominant splitting which we call {\em
hyperfine (HF)} splitting. The antiproton spin \SP\ leads to a further,
about two orders of magnitude smaller splitting called {\em
super-hyperfine (SHF)} structure, resulting in a quadruplet structure
(cf. Fig.~\ref{fig:HFS}).

The two-laser microwave triple resonance method \cite{Widmann:02}
allows to observe the transitions \nuHFp\ and \nuHFm\ within the
quadruplet that are associated with a flip of the {\em electron} spin.
The measured values agree well with the calculated ones by Korobov and
Bakalov \cite{Bakalov:98,Korobov:01},  but less with published values
of Yamanaka and Kino \cite{Yamanaka:00}. New results of Kino et al.
\cite{Kino:01}, however, are very close to our experimental results.
The experiment has therefore fully confirmed the presence of a
quadruplet structure originating from the hyperfine coupling of \Lp,
\Se, and \Sp, as predicted by Bakalov and Korobov \cite{Bakalov:98}.
The latest values of both theoretical groups agree with our result on
the level of the theoretical accuracy which is determined from omitting
order-$\alpha^2\approx 5\times 10^{-5}$ corrections from the
calculations. This again shows the impressive accuracy that the
variational three-body QED calculations are able to achieve. The
experimental accuracy is $\sim 3\times10^{-5}$, slightly smaller than
the theoretical uncertainty.

The excellent agreement between experiment and theory proves the
validity of the theoretical expressions of Bakalov and Korobov for the
HFS of \pbarhelium. The microwave resonance frequencies, \nuHFp\ and
\nuHFm,  are primarily related to the dominant \pbar\ orbital magnetic
moment.  On the other hand, the splitting between \nuHFp\ and \nuHFm\
is caused by the \pbar\ spin, and is directly proportional to the
spin-magnetic moment \mupbar\ of the antiproton. The observation of a
splitting in agreement with the theoretical value implies that the
spin-magnetic moment of proton and antiproton are equal within the
experimental error of 1.6~\%. This is consistent with an earlier (more
precise) determination of \mupbar\ from a fine structure measurement of
antiprotonic lead \cite{Kreissl:88}.

\section{Summary and Outlook}

\begin{figure}
        \begin{center}
        \epsfig{file=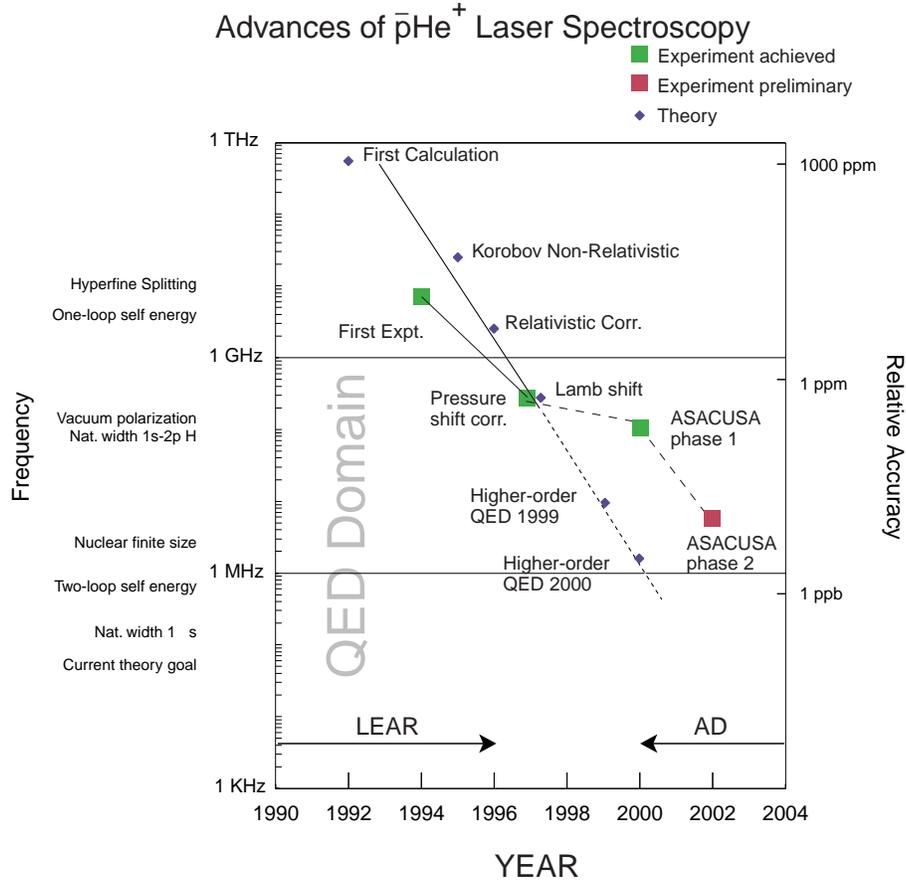,width=12cm}
        \end{center}
    \caption{\label{fig:advances} Evolution of both experimental and
        theoretical accuracy for laser spectroscopy of \pbarhelium.}

\end{figure}

Fig.~\ref{fig:advances} illustrates the development of the laser
spectroscopy measurements as well as the theoretical calculations over
the years. It shows a step-by-step progress, in which an advance of
experiments was followed by an improvement of the theoretical
calculations, which was repeated several times. The experimental
results provided the grounds for an advancement of three-body QED
calculations to a precision that has never been reached before for a
three-body system with two heavy centers (antiproton and helium
nucleus). The agreement between theory and experiment could be used to
perform a CPT test on the equivalence of proton and antiproton charge
and mass with an accuracy of 60 ppb, which will be improved by a factor
$3\ldots4$ in the near future. Making use of the ultra-low energy
antiproton beam provided by the ASACUSA RFQD, and two-photon
transitions with a new laser system currently under development, we
expect to increase the experimental precision by about another order of
magnitude in the next two years.

The first measurement of microwave-induced transitions within a state
in \pbhe proves its unusual hyperfine structure due to the large
angular moment of the antiproton. The agreement with theory provides a
measurement of the orbital angular moment of \pbar\ with an accuracy of
$6\times10^{-5}$. The value for the spin magnetic moment agrees with
earlier (more precise) measurements. An improvement of the accuracy of
one order of magnitude is necessary to improve the value for \mupbar,
which is known to only 0.3\%.

\clearpage\newpage

\newcommand{\SortNoop}[1]{} \newcommand{\OneLetter}[1]{#1}
  \newcommand{\SwapArgs}[2]{#2#1}

\end{document}